# Self Learning from Large Scale Code Corpus to Infer Structure of Method Invocations

1st Hung Phan
*Department of Computer Science*
*Iowa State University*
Ames, Iowa, USA 50010
hungphd@iastate.edu

*Abstract*—Automatically generating code from a textual description of method invocation confronts challenges. There were two current research directions for this problem. One direction focuses on considering a textual description of method invocations as a separate Natural Language query and do not consider the surrounding context of the code. Another direction takes advantage of a practical large scale code corpus for providing a Machine Translation model to generate code. However, this direction got very low accuracy. In this work, we tried to improve these drawbacks by proposing MethodInfoToCode, an approach that embeds context information and optimizes the ability of learning of original Phrase-based Statistical Machine Translation (PBMT) in NLP to infer implementation of method invocation given method name and other context information. We conduct an expression prediction models learned from 2.86 million method invocations from the practical data of high qualities corpus on Github that used 6 popular libraries: JDK, Android, GWT, Joda-Time, Hibernate, and Xstream. By the evaluation, we show that if the developers only write the method name of a method invocation in a body of a method, MethodInfoToCode can predict the generated expression correctly at 73% in F1 score.

## I. Introduction

Implementing method invocations are more challenges due to 2 reasons. First, developers need to remember a different sets of method Application Programming Interfaces (APIs) depending on their purpose. Second, the implementation of method invocation is also depending on the surrounding context of the code. These challenges cause the code developed by non-experience developers to be in the risks of having semantically incorrect. To alleviate the task of writing code, automatic approaches have been done for automatically assisting code from textual search by natural language description ( [1]–[4]). In these works, anyCode ( [4]) would rather be designed for searching the correct implementation from a sentence of NL text, instead of considering the input as a textual description about a correct implementation inside the code environment. The appearance of large scale repository hosting service such as Github ( [5]) brings a condition for a new direction of generating code which uses data-driven approach like ( [6]). However, this work also reveal the problem from reality code corpus, which contains noise causes the low accuracy in their code generation.

In this work, we want to overcome the challenges from prior works ( [4], [6]) by building a Phrase-based Statistical Machine Translation (PBMT) ( [7]) for learning information from actual large-scale code corpus. We deals with problems of noise data by two intuitions. First, source code elements such as method names are written in English Natural Language and they are usually mentioned in NL queries for code. Second, a suitable Machine Translation model for learning the implementation from the textual content of method invocation should embed the information of the surrounding code context instead of just the textual description of the method to the source and target language.

## II. Approach

We propose MethodInfoToCode, an approach that utilizes machine translation for inferring code from a textual description of the method invocation. MethodInfoToCode relies on practical data on large scale code corpus and embeds information surrounding code of each method invocation in the training data for the inference of implementation. We rely on the idea of building a parallel corpus which one side represents incomplete information developers can input to the code, while the other side represents the complete form for each method invocation and the surrounding context.

### A. Input by Developers

We consider that method name is the most important element we need to provide to developers. So, we allow developers to input textual description for method invocation by 3 parts. In the first part, developers can write the textual information about the method name. To help developers find the method name easily, we suggest the list of most textual similar method names from developers' textual input. The list of suggested method names is got from the database of method names we collected. In the second part, the developers can input the list of the variable they want to have in the implementation. In the last part, developers can write down the list of suggested words he or she wants to have. In general, developers have 2 forms of queries: query in compact form and query with details information.

TABLE I
INTRINSIC EVALUATION RESULT ON GITHUB PROJECTS

| | Intrinsic Evaluation with Configuration 1 | | | | | | | | |
|---|---|---|---|---|---|---|---|---|---|
| Library | Correct | Incorrect | OOSource | OOTarget | OOVoc | Total | Precision | Recall | F1-Score |
| GWT | 39635 | 22318 | 3082 | 28440 | 31522 | 93475 | 63.98% | 55.70% | 59.55% |
| Joda-Time | 27364 | 10608 | 51 | 1692 | 1743 | 39715 | 72.06% | 94.01% | 81.59% |
| JDK | 1053330 | 540997 | 3691 | 390626 | 394317 | 1988644 | 66.07% | 72.76% | 69.25% |
| Android | 471347 | 91753 | 5654 | 48662 | 54316 | 617416 | 83.71% | 89.67% | 86.58% |
| Hibernate | 53319 | 25305 | 4787 | 34090 | 38877 | 117501 | 67.82% | 57.83% | 62.43% |
| Xstream | 4671 | 1692 | 70 | 2949 | 3019 | 9382 | 73.41% | 60.74% | 66.48% |
| Total | 1649666 | 692673 | 17335 | 506459 | 523794 | 2866133 | 70.43% | 75.90% | 73.06% |

*1) Query in Compact Form:* In this case, developers only need to input the textual information relate to the method name. Consider the first example in Figure 1. The input provided by developers, `"get bit map"`, will be used for searching the most relevant method name in the corpus. After the list of most similar method names are suggested to developers, they will choose the method name they want to have. After the selection, all of the developers' input will be converted to a word `getBitMap#iden`. This input will be used as the source for translation.

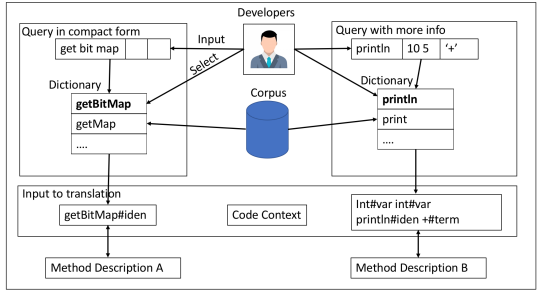

Fig. 1. Example of input by developers

*2) Query with Details Information:* Consider the other example shown in Figure 1. In this example, developers input `"println"` as a method name, 2 integers as variables, and `"+"` as a suggested word. Similar to the first example, all of the information inputted by developers will be encoded to input for the inference.

### B. Design the Source and Target Language for Translation

We consider the source language as incomplete representation about code, while target language is the complete representation about the code. We propose a parallel corpus with the consistent length between each pair of source and target language. Each token in the source language represents an incomplete form of the code, while each token in the target represents the full implementation and its roles inside each programming language elements. For method invocation, source side embeds information about method name, while the target side embed information as an expression which developers only add local variables, local methods or literal to get final code. For other AST Node types, we consider that the source language implies about the Partial Class Name (PQN) of each element, while the target language contains the Fully Qualified Name (FQN) for each element. We build a PBMT model for the inference from method names to expressions.

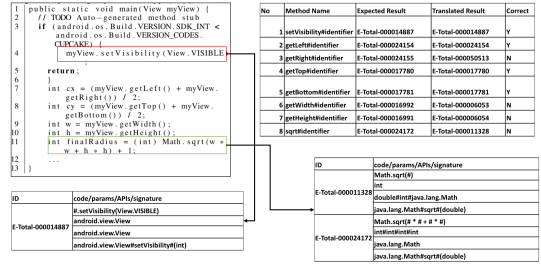

Fig. 2. Example of query in configuration 1 for code snippet in [8]

## III. EVALUATION

To do the evaluation, we select corpus from 1000 Github projects which frequently use APIs from 6 well-known libraries: Java Development Kit (JDK), Android, GWT, Joda-Time, Hibernate, and XStream. The input for translation contains only the method names along with tokens generated from other AST Node types in the source language for translation In total, there are 2.86 million of method invocations.The evaluation result for intrinsic data is shown in Table I. We show that if the developers only write method names, the F1 score is 73.06%, which is promising.

For illustration we collect an example from Program Creek in the post [8] is shown in Figure 2. We can see that if a developer writes `"set visibility"` and give the context since the developer wants to do some calculation relating to the position of the `View` object, the SMT model will automatically suggest an expression that set visibility to true of the correct receiver. Incorrect translated results often happen if developers query about a method name on complicate expression. In this example, if the developer inputs `"sqrt"`, the translated result is to accept a single integer variable, while the correct result requires four integers as parameters.